# Characterization of the CMS Endcap Timing Layer readout chip prototype with charge injection

**H. Sun,**[a,b,1] **D. Gong,**[b] **W. Zhang,**[a,b,1] **C. Edwards,**[c] **G. Huang,**[a] **X. Huang,**[a,b,1] **C. Liu,**[b] **T. Liu,**[b] **T. Liu,**[c] **J. Olsen,**[c] **Q. Sun,**[c,2] **J. Wu,**[c] **J. Ye,**[b] **L. Zhang,**[a,b,1]

[a] *Central China Normal University,*
*Wuhan, Hubei 430079, PR China*

[b] *Southern Methodist University,*
*Dallas, TX 75275, USA.*

[c] *Fermi National Accelerator Laboratory,*
*Batavia, IL 60510, USA.*

*E-mail*: qsun@fnal.gov

ABSTRACT: We present the characterization of a readout Application-Specific Integrated Circuit (ASIC) for the CMS Endcap Timing Layer (ETL) of the High-Luminosity LHC upgrade with charge injection. The ASIC, named ETROC and developed in a 65 nm CMOS technology, reads out a 16×16 pixel matrix of the Low-Gain Avalanche Detector (LGAD). The jitter contribution from ETROC is required to be below 40 ps to achieve the 50 ps overall time resolution per hit. The analog readout circuits in ETROC consist of the preamplifier and the discriminator. The preamplifier handles the LGAD charge signal with the most probable value of around 15 fC. The discriminator generates the digital pulse, which provides the Time-Of-Arrival (TOA, leading edge) and Time-Over-Threshold (TOT, pulse width) information. The prototype of ETROC (ETROC0) that implements a single channel of analog readout circuits has been evaluated with charge injection. The jitter of the analog readout circuits, measured from the discriminator's leading edge, is better than 16 ps for a charge larger than 15 fC with the sensor capacitance. The time walk resulting from different pulse heights can be corrected using the TOT measurement. The time resolution distribution has a standard deviation of 29 ps after the time-walk correction from the charge injection. At room temperature, the preamplifier's power consumption is measured to be 0.74 mW and 1.53 mW per pixel in the low- and high-power mode, respectively. The measured power consumption of the discriminator is 0.84 mW per pixel. With the ASIC alone or the LGAD sensor, The characterization performances fulfill the ETL's challenging requirements.

---

[1] Visiting scholars at SMU and performed this work at SMU.
[2] Corresponding author.

## 1. Introduction

A new detector, the CMS MIP Timing Detector (MTD), has been approved in the upgrade plan for the High-Luminosity Large Hadron Collider (HL-LHC) [1]. This detector will bring the ability to measure precisely the production time of Minimum Ionizing Particles (MIPs) for use in disentangling the approximately 200 nearly-simultaneous piled-up interactions in each bunch-crossing cycle (25 ns) of the HL-LHC. The MTD will also provide new capabilities for charged hadron identification and the search for long-lived particles. The MTD will consist of two sub-detectors, the Barrel Timing Layer detector (BTL) and the Endcap Timing Layer detector (ETL). Both sub-detectors will provide 30-40 ps resolution at the beginning of the HL-LHC operation and degrade eventually to 50-60 ps due to radiation damage by the end of the HL-LHC operation. The ETL will achieve this time resolution with two detector layers on each endcap, each layer resolving 40-50 ps per hit. The Low-Gain Avalanche Detectors (LGADs) [2-4] will be used in the ETL.

Building on the LGAD sensor studies' preliminary results, a dedicated ASIC, the Endcap Timing Read-Out Chip (ETROC), is being developed in a 65 nm CMOS technology. ETROC includes a 16×16 pixel matrix. Each pixel is 1.3×1.3 mm$^2$, the same size as the LGAD pixel. Each pixel in ETROC has a preamplifier, a discriminator, and a time-to-digital converter (TDC) for the Time-Of-Arrival (TOA) and Time-Over-Threshold (TOT) measurements. The TOT is used to correct the time walk due to the charge Landau distribution in LGAD [5, 6]. The power consumption of ETROC must stay below 1 W per chip, constrained by the system cooling capacity. This value translates to a power budget of 2.5 mW for the front-end analog readout circuits in each pixel. The time resolution of the ETL is determined by the LGAD sensor and ETROC together. The LGAD sensor has a jitter of about 30 ps due to non-uniform charge deposition. The ETROC contribution should be below 40 ps to achieve 50 ps overall time resolution per hit. The Most Probable Value (MPV) of the charge from the LGAD sensor is around 15 fC, and the expected operating range of the charges is 10-25 fC. Due to such a small signal from the LGAD sensor, the analog readout circuits, including the preamplifier and the discriminator, dominate electronics' jitter contribution and are critical for the CMS ETL precision timing performance.

The prototype of ETROC, called ETROC0, implements a single channel of the front-end analog readout circuits and has been developed and extensively tested under different conditions [7]. This paper comprehensively presents the characterization of ETROC0 by using an on-chip charge injector and focuses on the performance study of ETROC0 itself using charge injection without a real LGAD signal involved. Two cases, with a soldered capacitance or the LGAD sensor seen as a capacitance, are used in the characterization and discussed in Section 3. Note that the LGAD sensor could still be involved and biased in the latter case. More comprehensive tests of ETROC0, including the TID test at CERN and the beam test at Fermilab to evaluate its performance with LGAD pulses, can be found in the reference [7], and beyond the scope of this paper.

## 2. Circuit Design

### 2.1 Architecture of pixel readout

Figure 1 illustrates the block diagram of the readout circuits of a single-pixel in ETROC. The LGAD sensor, reversely biased by a negative high voltage (HV), is DC coupled to the analog readout circuits via a bump pad. The analog readout circuits include a preamplifier, a discriminator, and a voltage Digital-to-Analog Converter (DAC). The preamplifier is a Trans-Impedance Amplifier (TIA), which converts the input current signal to a voltage signal. The discriminator compares the analog voltage pulse with a threshold and outputs a digital pulse. The DAC sets the threshold of the discriminator. We can observe the output signals of the preamplifier and the discriminator with an analog buffer and a digital buffer, respectively. Each pixel contains a charge injection circuit for testability. The digital readout parts such as a TDC [8] and a Random-Access-Memory (RAM) are integrated with the analog readout circuits in the second prototype ASIC called ETROC1, which is beyond the scope of this paper.

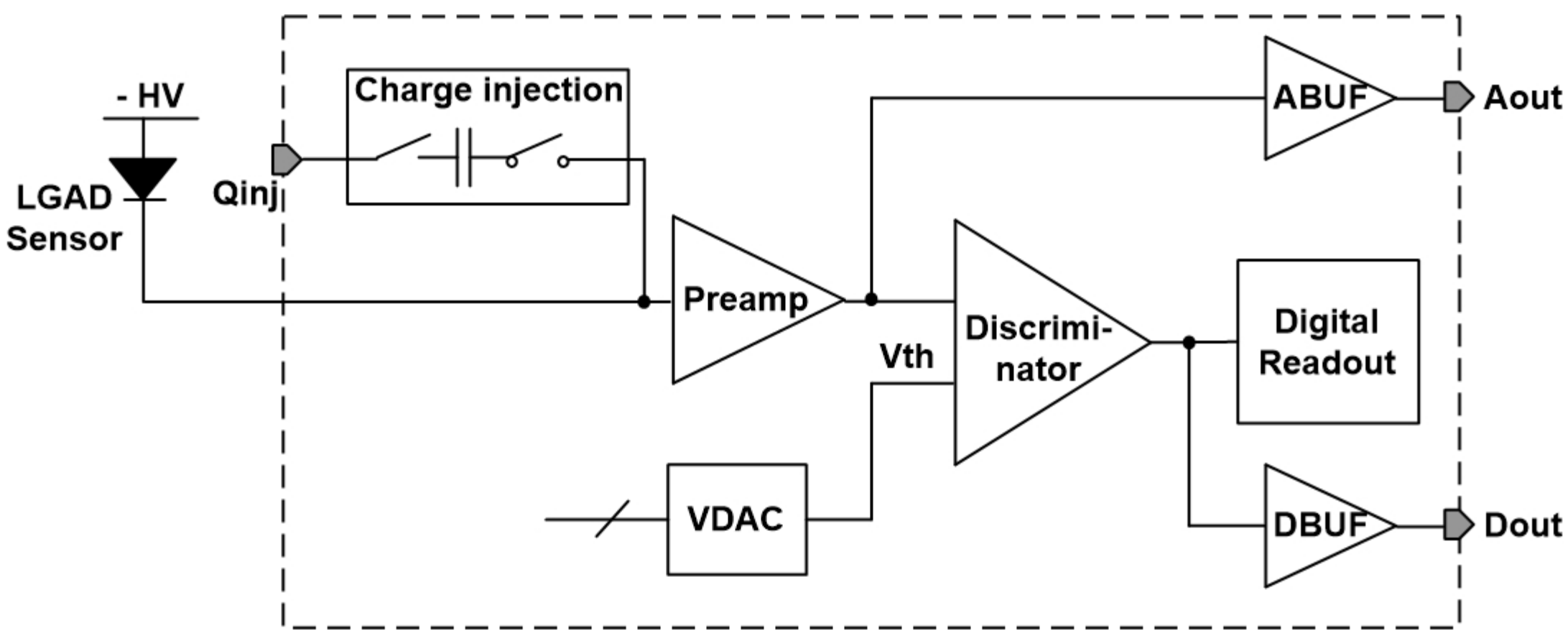

**Figure 1.** Block diagram of the analog readout circuits of a single pixel.

## 2.2 Preamplifier

The preamplifier's topology in ETROC is based on a timing-measurement-optimized TIA, which has been well studied and carried out [9-12]. The schematic of the preamplifier is shown in figure 2(a). The detector can be modeled as a current source $I_{in}$ flowing from the TIA to the reversely-biased LGAD sensor. The corresponding input charge ($Q_{in}$) is equal to $I_{in} \cdot T_d$, where $T_d$ is the current's effective duration. Through the total detector capacitor $Cs$, the current $I_{in}$ is converted into a negative voltage pulse with the amplitude:

$$V_{in} = \int (I_{in}(t)/Cs)\, dt = Q_{in}/Cs, \tag{1}$$

where the total capacitance $Cs$ include the sensor capacitance and the parasitic capacitance on the preamplifier's input node.

Jitter is one of the contributions to the total time resolution and could be approximately calculated as follow:

$$\sigma_{jitter} = \frac{V_n}{\frac{dV}{dt}}, \tag{3}$$

where $V_n$ is the preamplifier output noise, and $\frac{dV}{dt}$ is the signal slew rate. The output noise and the slew rate are then given by:

$$V_n \propto G_{PA} \times \sqrt{BW} \propto \frac{G_{PA}}{\sqrt{t_r}} \tag{4}$$

and

$$\frac{dV}{dt} = \frac{G_{PA} \times V_{in}}{dt} = \frac{G_{PA} \times Q_{in}}{Cs\sqrt{t_r^2 + t_s^2}}, \tag{5}$$

where $G_{PA}$ is the preamplifier gain, $BW$ is the preamplifier bandwidth, $t_r$ is the rise time of the preamplifier output signal, $V_{in}$ is the preamplifier input voltage as defined in equation (1), and $t_s$ here is the rise time of an LGAD signal. The jitter can be expressed by combining the equations above as:

$$\sigma_{jitter} \propto \frac{Cs}{Q_{in}} \sqrt{\frac{t_r^2 + t_s^2}{t_r}}. \tag{6}$$

For a given $t_s$, matching $t_r$ with $t_s$ can minimize the timing jitter. We can optimize the jitter with a small sensor capacitance, a large input charge, a faster rise time of the LGAD signal and preamplifier.

The preamplifier consists of two stages: a cascade amplifier (M1 and M2) as the first stage and a source follower (M3) as the second stage. The bias current of the input transistor (M1) has two components: the constant current $I_{B1}$ of 0.16 mA and a tunable current $I_{B2}$ from 0.25 mA to 0.95 mA. The transistor M2 and its gate voltage Vb set the DC operating point of M1. Vb is the replica bias voltage from $I_{B1}$. The feedback resistor $R_f$ is programmable to adjust the gain of the preamplifier. The load capacitance $C_L$ of the first stage is also programmable to optimize the bandwidth. The drain current $I_{B3}$ of M3 is within 0.15 mA and generated by a resistor. The gain of the first stage and the second stage are negative and positive, respectively. Since the preamplifier's input signal is a negative pulse, the output signal is a positive pulse, whose rise edge is the leading edge and falling edge the trailing edge.

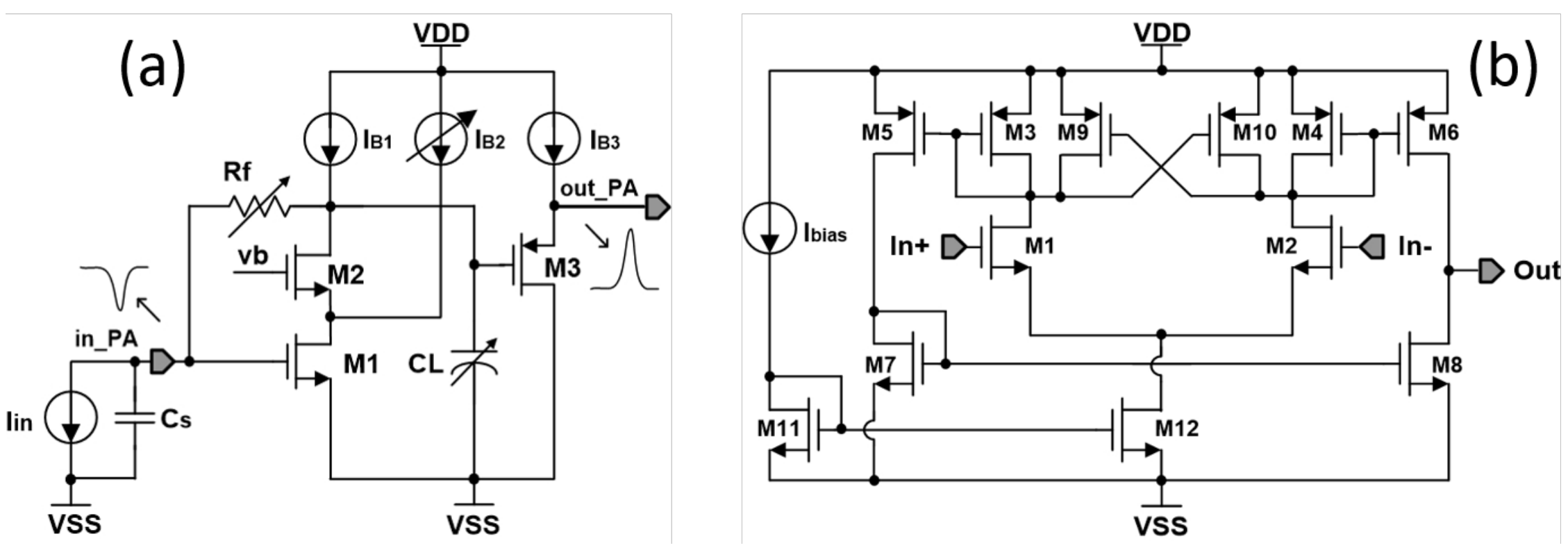

**Figure 2.** Schematic of the preamplifier (a) and the comparator (b).

We consider both the leading and trailing edges in the design of the preamplifier. A faster leading edge can be achieved with a higher bias current and a smaller load capacitor. When the bias current decreases from its highest to its lowest value, the preamplifier output's leading edge time can be set from 700 to 800 ps with a bandwidth from 270 to 170 MHz. When the bias current is the lowest, the additional load capacitance $C_L$ can also be variable from 0 to 160 fF to adjust the leading edge time from 800 to 960 ps. Small load capacitance leads to a fast leading edge, but introduces more noise because of the large bandwidth. Therefore, the load capacitance can be selected to fine-tune the jitter. The default setting is to use the smallest load capacitance.

The trailing edge time of the preamplifier output depends on the time constant, the product of the total capacitance $Cs$ and the input impedance $R_{in}$. The input impedance $R_{in}$ is given by the feedback resistance $R_f$ divided by the open-loop gain. The open-loop gain depends on the bias current. Thus, $R_f$ is programmable with four settings (4.4, 5.7, 10, and 20 kΩ) to adjust the gain (from 3.4 to 15.7 kΩ) and the trailing edge time (from 1.8 to 4 ns) of the preamplifier output. Based on the simulation with LGAD signals, the default feedback resistor of 5.7 kΩ is selected.

The bias current $I_{B2}$ allows different trade-offs between the power consumption and the timing performance. A larger signal slew rate ($\frac{dV}{dt}$) and a faster rise time of the preamplifier output $t_r$ can be achieved at a higher bias current, and thus the better performance is expected. The lowest bias current is the default setting, representing the low-power mode of the preamplifier. The preamplifier in the high-power mode operates with the maximum value of $I_{B2}$.

### 2.3 Discriminator and DAC

The discriminator consists of three stages of fully differential amplifiers, a comparator, and an internal buffer. The three stages of amplifiers receive the small input pulses, which are as low as a few millivolts in some extreme cases, and generate the larger output pulses comfortable for the comparator. The overall gain for three stages of amplifiers is 35 dB with a bandwidth of around 390 MHz. The comparator discriminates the differential input at the crossing point with programmable hysteresis. The internal buffer delivers the digital output to the following circuits. The internal buffer is composed of two CMOS inverters and relieves the loading pressure.

We will focus on the comparator's design since the three stages of amplifiers and the internal buffer are relatively simple. As shown in figure 2(b), the comparator has two stages: The first stage is a specialized and high-gain common-source amplifier constituted by the MOSFET pair

M1 and M2. The transistor pair M3 and M4 are the PMOS diode-connected loads of the amplifier. M11 and M12 provide the first stage's bias current (160 μA) with a source current $I_{bias}$ of 20 μA, shared with the three-stage amplifiers. The high-gain differential stage digitizes the differential input at the crossing point with a tunable hysteresis generated by the transistors M9 and M10. The hysteresis ranging from 0 (by default) to 1 mV is used to alleviate ringing due to noise. The second stage converts the differential output of the first stage into single-ended and provides additional gain. The leading edge of the discriminator pulse provides the TOA, while the trailing edge, combined with TOA, provides the TOT.

The discriminator threshold (Vth) is connected to the inverting input and set by an internal 10-bit resistor string DAC [13]. Since the preamplifier's baseline varies with temperature, bias setting, and irradiation, the DAC output ranges from 0.6 to 1 V with an LSB (Least significant bit) of 0.4 mV. A reference voltage generator provides a 1 V reference voltage to the DAC. A first-order RC filter is designed so that the threshold voltage's noise is well below 0.16 mV to minimize the DAC noise contribution to the timing performance.

### 2.4 Charge injector

The charge injection circuit (charge injector) allows a simple way to check and characterize the front-end readout circuits in ETROC. Figure 3 presents the schematic of the charge injection circuit. Qinj_P/N is a differential pulse signal in CERN Low-Power Signaling (CLPS) standard [14]. The electrical port (eRx) is a differential receiver implanted from the low-power Giga-Bit Transceiver (lpGBT) [15, 16]. Qinj is a fast voltage step pulse on the gate of NMOS switch M1 with the power supply voltage amplitude. Qinj generates a voltage step, QV. The amplitude of QV depends on the bias current, $I_{bias}$, which is programmable by a 5-bit current DAC in the range of 1 to 32 μA. QV ranges from 10 to 320 mV and corresponds to the injected charge of 1 to 32 fC. The following equation gives the injected charge $Q_{in}$:

$$Q_{in} = V_{QV} \cdot C_q = I_{bias} \cdot R_q \cdot C_q, \tag{7}$$

where $R_q$ = 10 kΩ, and $C_q$ = 100 fF. Both $R_q$ and $C_q$ are integrated in each pixel.

In ETROC0, the node of QV is accessible off-chip. We can measure the output voltage on this pad when using Qinj as the input step or directly apply a fast voltage step pulse.

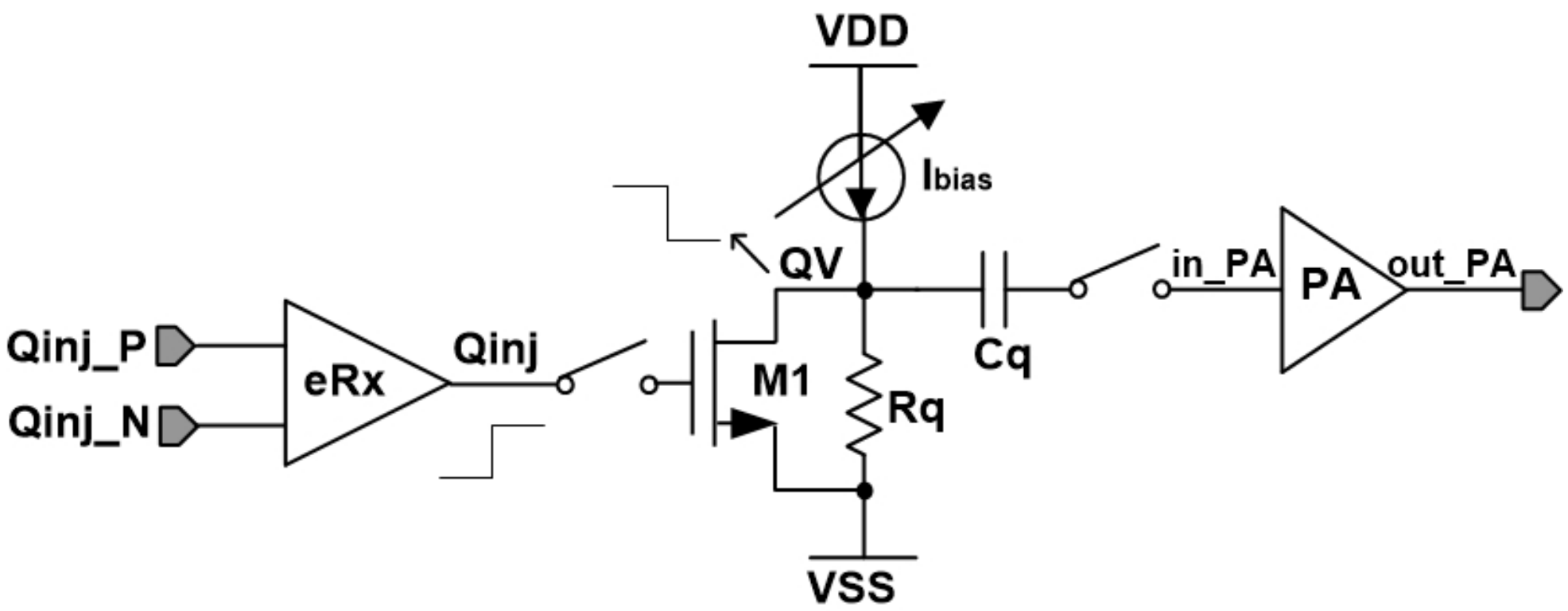

**Figure 3.** Schematic of the charge injector.

## 3. Measurements

The performance of ETROC0 has been characterized by charge injection. In this section, we present the test bench setup and the characterization results.

### 3.1 Test bench setup

We developed two sets of the printed circuit boards (PCBs) to carry out the tests with charge injection. One set is used for the ASIC tests alone, and the other one for the module, consisting of the ASIC and the LGAD sensors.

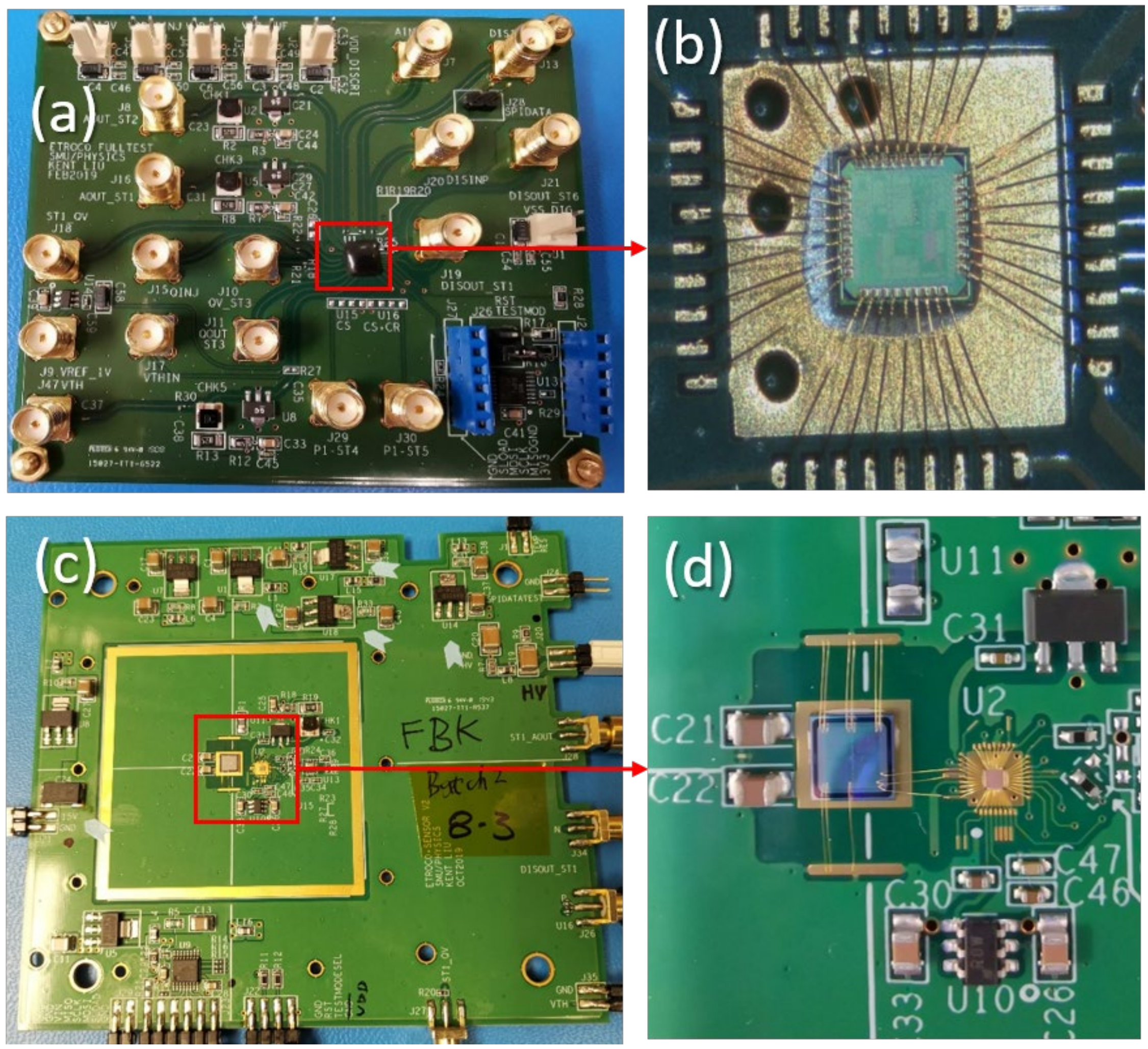

**Figure 4.** Photographs of a dedicated test board for ASIC alone (a), the ETROC0 die wire bonded to the test board (b), a sensor test board (c), and zoom of figure (c) where an ETROC0 chip is wire bonded to a 2×2 LGAD sensor (d).

A photograph of the dedicated board for the ASIC alone is shown in figure 4(a). Figure 4(b) illustrates how an ETROC0 die was wire-bonded on the dedicated board. An external capacitor is soldered on the preamplifier's input, together with about 1.5 pF parasitic capacitance on the board, to emulate the equivalent load of the LGAD sensor.

Figure 4(c) shows the module test board. The sensors used were 2×2 FBK [17,18] or HPK [5] LGAD sensor array with a 50 µm thickness and 1.3×1.3 mm$^2$ pixel size. Since ETROC0 consists of a single channel, only one pixel of the LGAD sensor was wire bonded to a stepping-stone pad on the PCB and connected to the preamplifier's input, as seen in figure 4(d). The LGAD sensor was glued onto a square High Voltage (HV) pad, which allows applying an external and varied HV to the LGAD to investigate the module performance. The typical HV value is –300 V to ensure that the FBK LGAD sensor maintains a gain of 10 and has a limited and stable sensor capacitance.

During the tests, a pulse pattern generator (Tektronix Model DTG5274) provides a sharp step pulse with a signal rate of 1 MHz. The rise time of the pulse is measured to be 125 ps (20%-80%). The step pulse can be injected into Qinj_P/N or QV input pins via Sub-Miniature version A (SMA) connectors. The step pulse goes through an on-chip capacitor of 100 fF and produces the injected charge at the preamplifier's input. Note that the on-chip capacitance has a variation of ±15% due to the process corner variation.

ETROC0 has an on-chip analog buffer to monitor the waveform of the preamplifier. The gain of the analog buffer is about 0.7 in the post-layout simulation. The test board of ETROC0 has an on-board RF amplifier (Mini-Circuits GALA-S66+) with a gain of about -10 (inverted) and a bandwidth of about 2 GHz. The preamplifier output (after the on-board inverting amplifier) and the discriminator output are accessible on SMA connectors.

A real-time oscilloscope (Tektronix Model DSA71254B) with a sampling rate of 50 Giga-samples/s and an analog bandwidth of 12.5 GHz is used to capture and save the output waveforms of the preamplifier and the discriminator. The trigger signal of the oscilloscope comes from the same source as the charge injection pulse.

### 3.2 Output waveforms

Figure 5(a) compares the preamplifier's output waveforms for different values of the soldered capacitor ($C_{sold}$) with an injected charge of 15 fC. The waveforms include the effects of the on-chip analog buffer and the on-board amplifier. The typical gain of the on-board amplifier is -10; thus, the preamplifier's observed pulses are negative. The amplitude of the preamplifier output decreases with the soldered capacitance, the time of the peak appears later, and the trailing edge also becomes slower.

The discriminator's output waveforms for different values of the soldered capacitor with an injected charge of 15 fC are shown in figure 5(b). A proper threshold for the discriminator is essential to get the best time resolution. The optimal threshold is usually a few millivolts above the preamplifier baseline. Since the discriminator output is accessible off the chip, we conduct the DAC scan searching for the baseline, and it is determined by the peak of the noise rate [19] of the discriminator output. A threshold voltage of 3.5 mV above the baseline is applied throughout the tests. The threshold voltage should be calibrated and reset before each run of the tests because the baseline varies slowly with temperature, bias setting, and the reverse High Voltage of the LGAD sensors. The discriminator's digital buffer has a 50-Ω termination on the oscilloscope; thus, the probe signal's amplitude halves. As shown in figure 5(b), the pulse width (known as TOT) increases with the soldered capacitance, which is consistent with the trailing edge of the preamplifier in figure 5(a).

With the sensor capacitance, the preamplifier probe's waveforms for different injected charges, both in low-power and high-power modes, are shown in figure 5(c). The larger the injected charge, the larger is the amplitude of the preamplifier output, while the peak time is the same. For the same injected charge, a sharper pulse on the preamplifier output can be observed in the high-power mode with a larger slew rate and a faster leading edge.

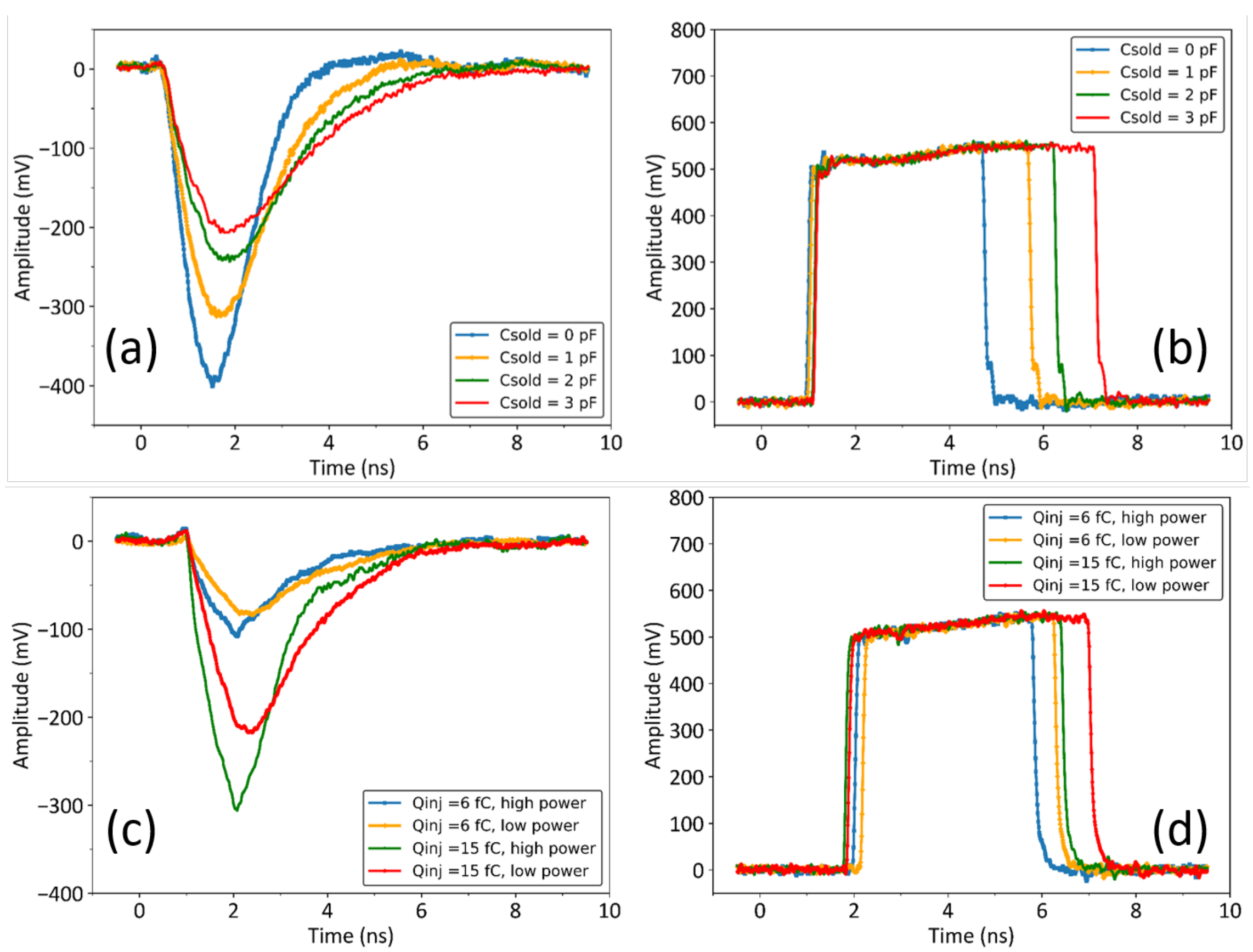

**Figure 5.** Output waveforms of the preamplifier (a) and discriminator (b) for different soldered capacitance when Qinj = 15 fC, and output waveforms of the preamplifier (c) and discriminator (d) for different injected charges in two power modes with the sensor capacitance.

Figure 5(d) presents the discriminator output waveforms corresponding to the preamplifier pulses in figure 5(c). The pulse width of the discriminator output increases with the injected charge but decreases with the power consumption. The pulse leading edge becomes earlier for a larger charge and with higher power consumption.

The time separation between two bunches at the HL-LHC is 25 ns. The nominal hit occupancy for each pixel in ETROC is 1%, and therefore it translates to a signal rate of 400 kHz. With an input signal rate of 1 MHz, the output waveforms are generated by the rising edge of the input signal (Qinj), as shown in figure 5. Those waveforms indicate that the analog processing of the input signals is certainly accomplished within 25 ns. Thus ETROC0 is completely ready for new hits at the next bunch crossing. The trailing edge of Qinj will not disturb the front-end analog processing and the timing measurements.

### 3.3 Charge injection property

For a fixed charge injection capacitor, the injected charge is determined by the voltage amplitude of the step pulse (QV). In the charge injection circuit, the QV voltage is linear to the bias current, which is programmable via the 5-bit current DAC. Figure 6(a) compares the ideal and measured values of the QV voltage with the injected charges (i.e., the preset DAC values) from 1 to 32 fC.

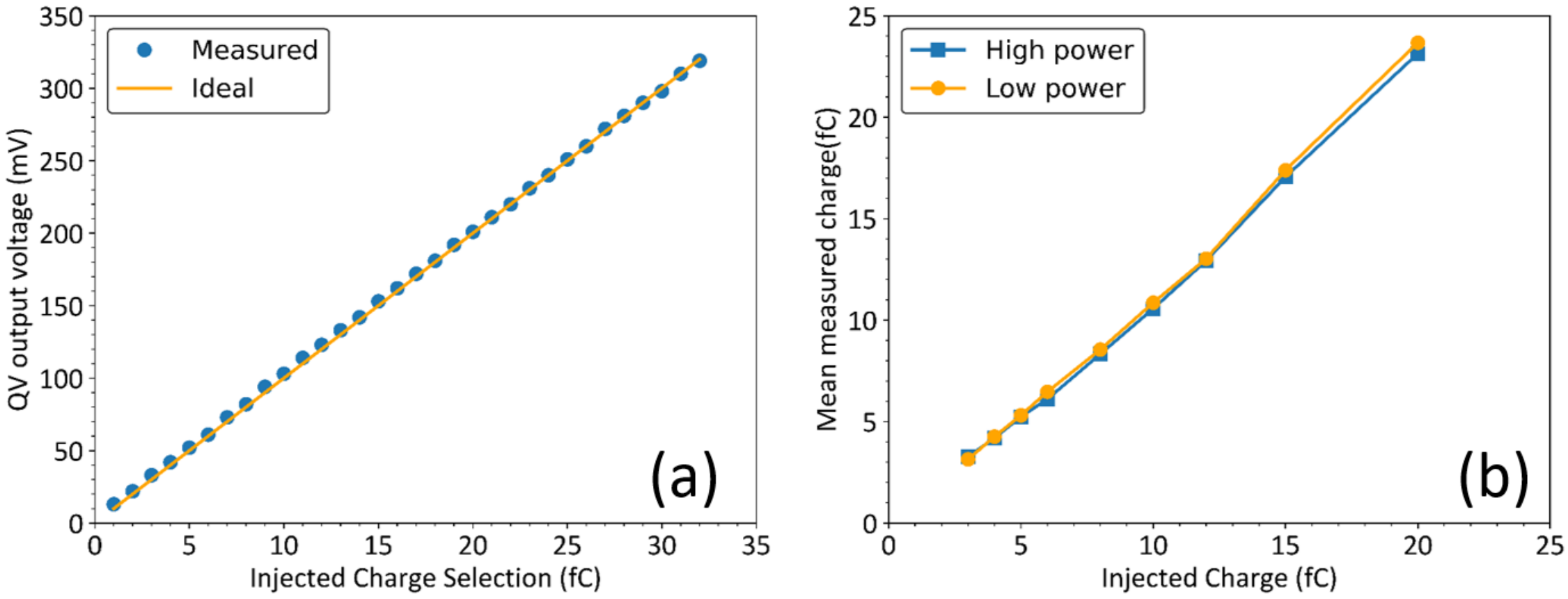

**Figure 6.** QV voltage (a) and measured charges (b) of the charge injection for different injected charges.

The measured QV voltage has a linear relationship with the preset DAC values, as expected. The difference between the measured and ideal values is below 10% and within the process corner and temperature variations.

Figure 6(b) presents the average calculated input charge in the low-power and the high-power modes with the sensor capacitance. The preamplifier's input charge can be derived by integrating the preamplifier output waveform and then dividing by the trans-impedance. All the post stages' trans-impedance and gain are required to derive the input charge: the trans-impedance of the first stage of the preamplifier is estimated as 4.4 kΩ from the simulation. The gain of the internal analog buffer and the on-board amplifier is 0.7 (the typical value) and 10 (the absolute value), respectively. The calculated charge in different power modes matches the injected charge within 20%. The difference is similar to the gain variations, which are measured to be as large as ±20% based on the standalone tests of the internal analog buffer and the on-board amplifier.

### 3.4 Preamplifier performance

Figure 7(a) compares the amplitude of the preamplifier output versus the injected charge in different power modes on two sets of test boards: (1) a 2 pF capacitor is soldered on the board with ETROC0 alone; (2) the LGAD sensor is wire bonded to ETROC0 and seen as the sensor capacitance. In both cases, the preamplifier is configured with the default feedback resistance and the default load capacitance. The larger the injected charge, the larger is the pulse amplitude. The amplitude in the high-power mode (blue lines) is larger than that in the low-power mode (orange lines) because the preamplifier's gain increases with the bias current. Compared to the ASIC results with a soldered capacitance of 2 pF, the pulse amplitude measured from the module (ASIC + LGAD sensor) is smaller. This result indicates that the real sensor capacitance is larger than a total capacitance of 3.5 pF (2 pF soldered capacitance plus 1.5 pF parasitic capacitance).

The preamplifier jitter was measured by using the histogram measurement method embedded in the oscilloscope. The window is below the on-board amplifier's baseline with a voltage threshold equivalent to 1.3 fC charge, noting that the on-board amplifier's waveform is a negative pulse. Figure 7(b) demonstrates the jitter of the preamplifier output as a function of the injected charge with a soldered capacitance of 2 pF and the sensor capacitance. All the measured jitter approximately inversely proportional to the injected charge, following equation (6). As expected, lower jitter is achieved with higher power consumption. The jitter is a bit higher with the sensor capacitance. In the low-power mode, the leading-edge jitter for the injected charge larger than 15 fC is below 10 ps with the soldered capacitance and within 15 ps with the sensor capacitance, respectively. The measured leading-edge jitter in both cases is better than the design requirements.

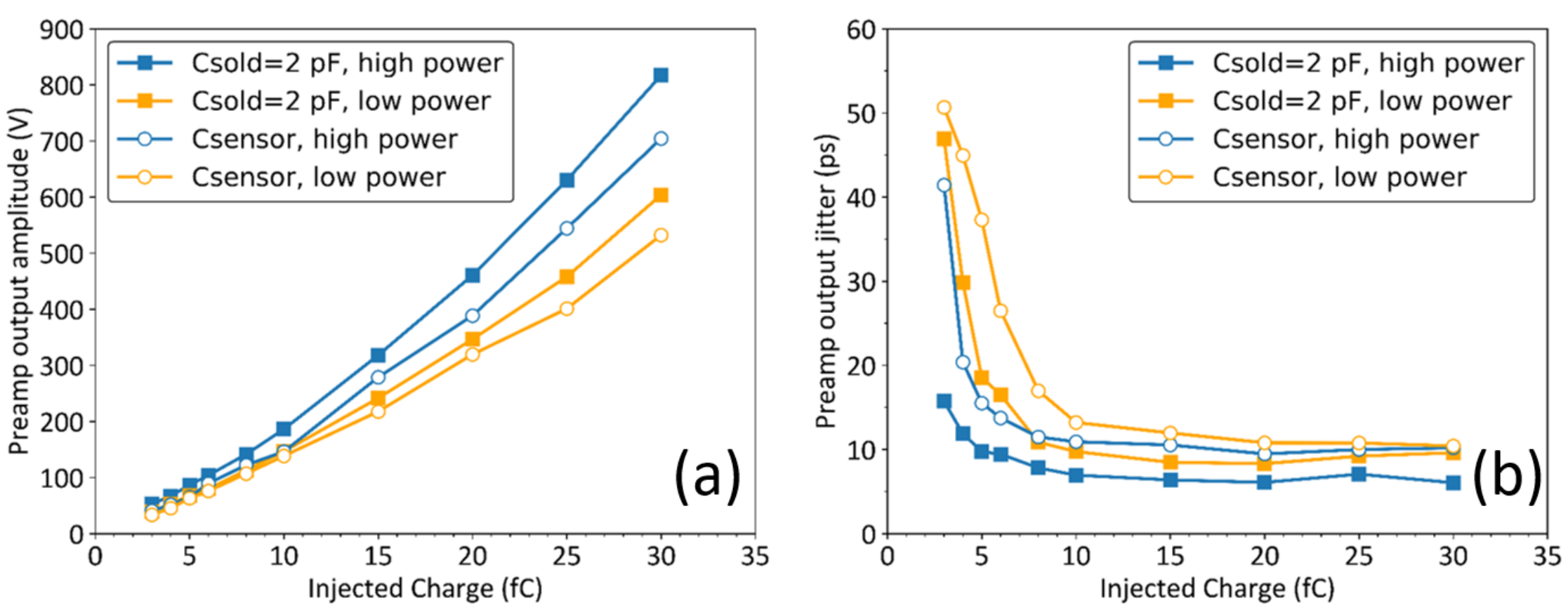

**Figure 7.** Amplitude (a) and leading-edge jitter (b) of the preamplifier for different injected charges.

### 3.5 Discriminator performance

Figure 8(a) demonstrates the discriminator TOT as a function of the injected charge for a discriminator threshold of 1.3 fC (or 3.5 mV). The measured discriminator TOT reaches a plateau of 5.5 ns and 5 ns with high charges in the low-power and the high-power modes, respectively.

For the jitter measurements, the time was measured at half of the maximum amplitude. Figure 8(b) compares the measured jitter at the leading edge of the discriminator output in different power modes and with different total capacitance. It reaches a plateau with high charges in both modes. In the sensor capacitance case, the leading-edge jitter in low-power mode is better than 16 ps when injected charges are more extensive than 15 fC. For the ASIC test alone with a soldered capacitance of 2 pF, the jitter performances are even better. The higher power consumption buys smaller jitter for all charge points. The jitter as a function of the injected charge in different cases fits the equation's theoretical prediction (6).

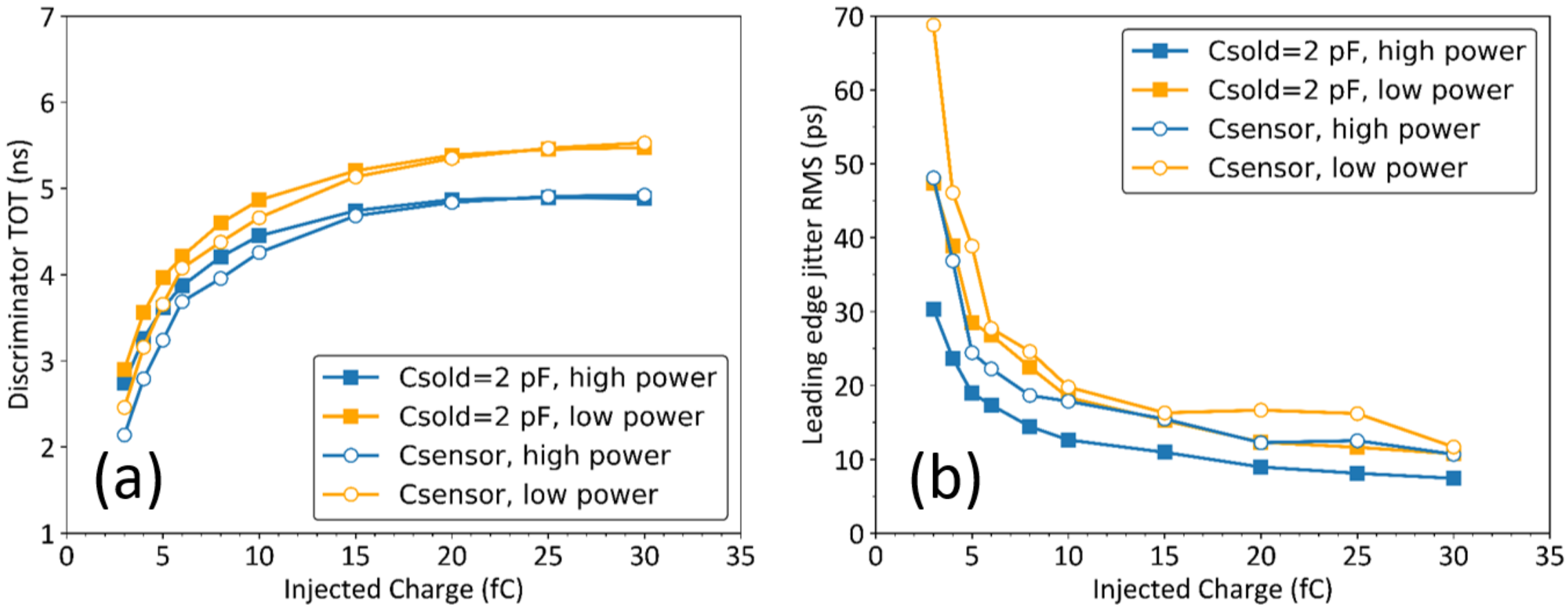

**Figure 8.** TOT (a) and leading-edge jitter (b) of the discriminator for different injected charges.

Jitter as a function of the total capacitance is shown in figure 9(a), with the injected charges of 6 fC and 15 fC. The linear relationship is observed as expected. For a small charge, the linear function's slope as expressed in equation (6) is larger — the smaller the total capacitance, the more benefit the jitter gains.

Figure 9(b) shows the leading-edge jitter derived from the discriminator at different reverse bias voltage on the sensor board where an FBK LGAD sensor is wire bonded to ASIC and seen as the sensor capacitance. The injected charge is 15 fC. No significant dependence is observed on the reverse HV from 240V to 360 V, indicating that the FBK LGAD sensor's capacitance reaches a plateau in this range. Note that the leakage current of the modules was measured to be below 3 μA. Similar consistency is also observed on a few other module boards that used HPK LGAD sensors.

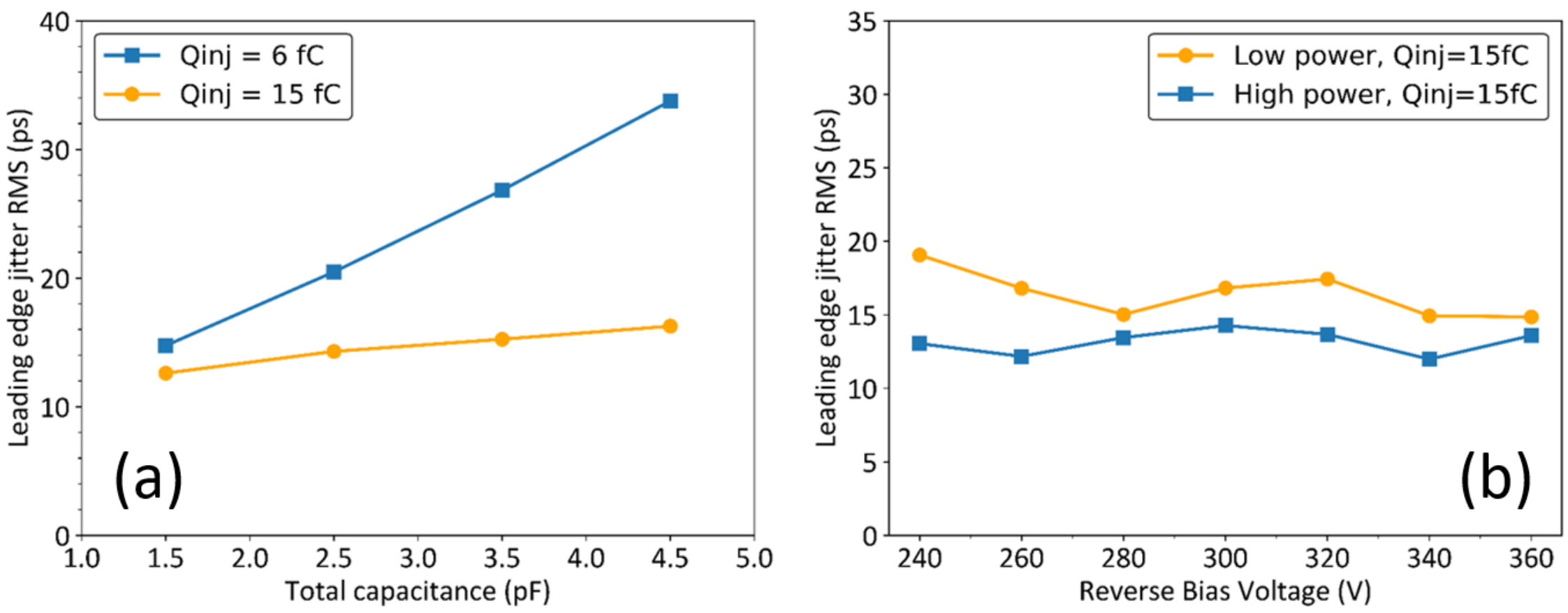

**Figure 9.** Leading-edge jitter of the discriminator as a function of the total capacitance (a) and discriminator output jitter at different reverse bias voltages when Qinj = 15 fC (b).

### 3.6 Time-walk correction

The time walk is the dependence of the threshold-crossing time on the pulse height. The ETL adopts the discriminator TOT as an estimate of the input charge [1]. The time walk is corrected by measuring the TOT of the discriminator output and removing the TOT's dependence from the TOA. The jitter from the preamplifier and the discriminator has to be kept below 40 ps [1].

On two sets of the test boards shown in figure 4, we use the discriminator TOT for the time-walk correction. Charged injection with 3, 4, 5, 6, 8, 10, 12, 15, 20, and 30 fC are applied to create time walk artificially. In the sensor board's measurement, the high voltage (HV) is fixed at –300 V. The preamplifier is set with the default gain and in the low-power mode, and the threshold (Vth) is placed at 3.5 mV above the preamplifier baseline. Figure 10(a) shows the raw data of the TOA, which is fitted with a 3rd-order polynomial,

$$TOA_{fitted} = a + b \times TOT + c \times TOT^2 + d \times TOT^3. \tag{8}$$

The orange line corresponds to this fit with four constraints that are applied in the correction. A TOA spread due to a time walk of about 1 ns is observed. The corrected TOA, which are the differences between the raw data and the fitted ones, are presented in figure 10(b). Figure 10(c) shows the average TOA residuals versus the average TOT after time-walk correction. The calculated residuals are in a peak-to-peak range of 15 ps, meeting the design requirements of ETROC. Figure 10(d) illustrates the time resolution, the longitudinal projection of the corrected

TOA shown in figure 10(b). The time resolution distribution has a standard deviation of 29 ps below the required 40 ps time resolution of ETROC contribution.

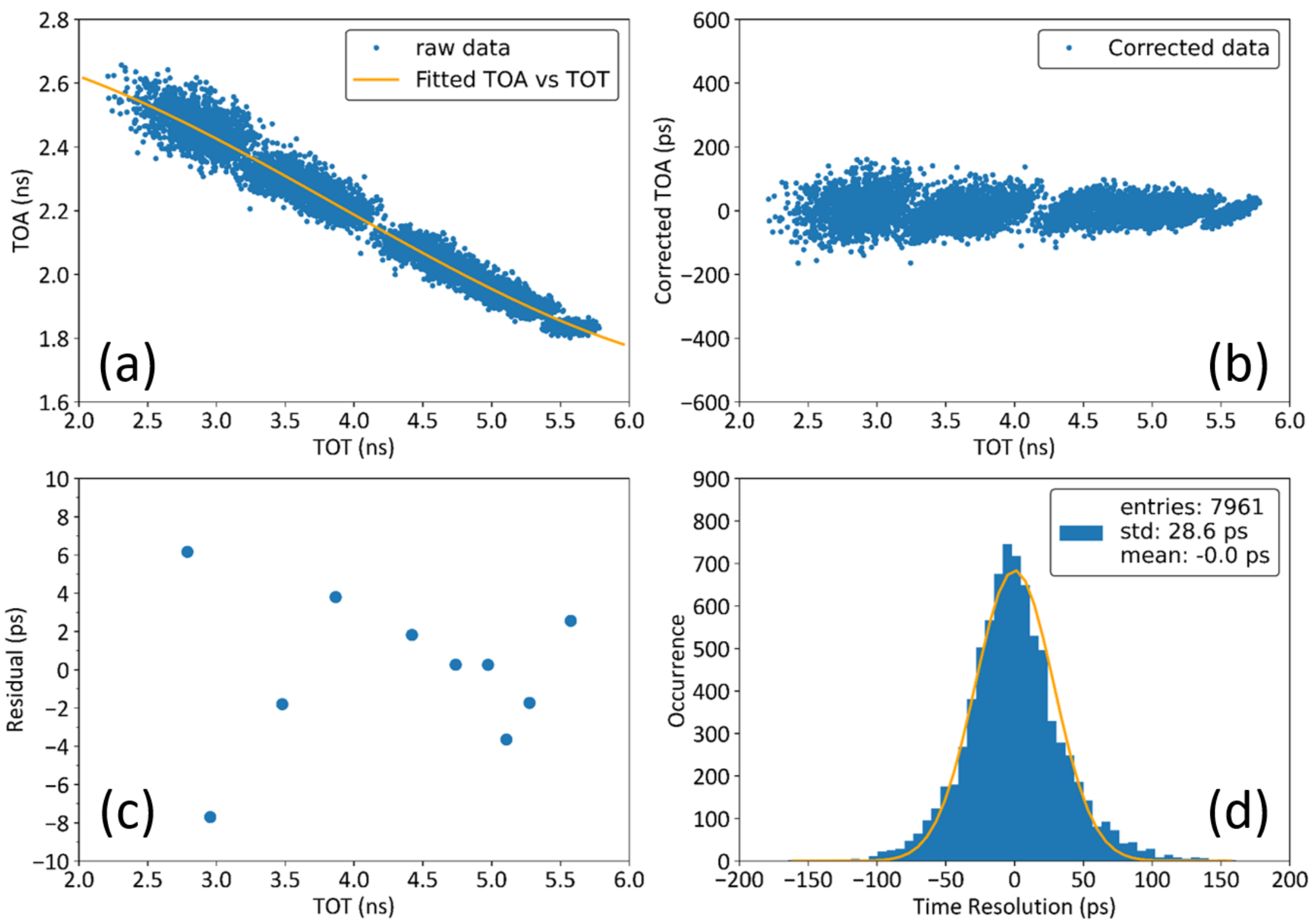

**Figure 10.** Time-walk correction (TOA vs. TOT). (a) Raw data. (b) Corrected data. (c) Average Time-walk residuals. (d) Histogram of time resolution (corrected TOA).

Table 1 summarizes the time walk correction conducted in different testing cases. For the expected charge range in the ETL operation, the standard deviation of the time resolution is within 25 ps. For a wider range with small charges, the time resolution is still no more than 30 ps. The time performances of the ASIC with a soldered capacitance of 2 pF and the sensor capacitance are comparative. ETROC0 can achieve better time performance with higher power consumption. All the results meet the design requirements of the ETROC.

**Table 1.** Time walk correction in different test conditions

| Charge range | Test boards | Time resolution (ps) | |
|---|---|---|---|
| | | Low-power mode | High-power mode |
| 3-30 fC | ASIC with $C_{sold}$ =2 pF | 30 | 20 |
| | Module with sensor capacitance | 29 | 21 |
| 10-30 fC | ASIC with $C_{sold}$ =2 pF | 24 | 15 |
| | Module with sensor capacitance | 21 | 15 |

### 3.7 Power consumption

The power consumption of the preamplifier and the discriminator is summarized in table 2. At room temperature (25 °C), the preamplifier's measured power consumption is 0.74 mW and 1.53 mW in the low-power and the high-power mode, respectively, with the default feedback resistance and the default load capacitance. The power consumption of the discriminator is measured to be 0.84 mW. The measured power consumptions agree well with the simulated values and fulfill the power consumption requirements. At -20 °C, where the ASIC will operate in the future application, the preamplifier's power consumption is estimated to be 0.70 mW and 1.55 mW in the low-power and the high-power mode, respectively, from the post-layout simulation. The simulated power consumption of the discriminator at low temperature is 0.75 mW. The power consumption performance of the ASIC will be studied by using a climate chamber in the future.

**Table 2.** Power consumption (mW) of the preamplifier and the discriminator.

| Operating temperature | Condition | Preamplifier | | Discriminator |
|---|---|---|---|---|
| | | Low-power mode | High-power mode | |
| 25 ℃ | Measured | 0.74 | 1.53 | 0.84 |
| | Simulated | 0.76 | 1.59 | 0.87 |
| -20 ℃ | Simulated | 0.70 | 1.55 | 0.75 |

## 4. Conclusion and outlook

This paper presents the characterization of the analog front-end circuits for CMS ETL HL-LHC upgrade. The design of the preamplifier and the discriminator is depicted. A charge injection circuit is implemented to evaluate the performance of ETROC0 without the real LGAD signal. The jitter contribution of the analog front-end circuits to the time resolution, derived from the leading edge of the discriminator output, is within 16 ps for a charge larger than 15 fC, either with a soldered capacitance or with the LGAD sensor seen as a capacitance. After time walk correction in all ranges of the charge injection, the residual is less than 15 ps (peak-to-peak), and the standard deviation of the time resolution distribution is 29 ps. The preamplifier's power consumption is measured to be 0.74 mW (low-power mode) and 1.53 mW (high-power mode) per pixel at room temperature. The power consumption of the discriminator is 0.84 mW per pixel at room temperature. The measured performance of the ASICs fulfills the challenging requirements.

ETROC0 has been tested at Fermilab in a proton beam to evaluate the module's performance with LGAD pulses [7]. ETROC1 has a 4×4 pixel array of the readout circuits, including the same front-end design as ETROC0 and TDC for time measurements. The bench test and test beam measurements of ETROC1 are ongoing. More irradiation tests will follow.

The next iteration of the ASIC, ETROC2, will contain the full functionality of the ETROC by using the full 16×16 pixel matrix. The ETROC2 design builds on experience gained through the development and the tests of ETROC0 and ETROC1. The remaining core functionalities, including the on-chip automatic threshold calibration, the slow and fast control interfaces, the Phase-Locked Loop (PLL), and the digital readout logic, are being developed in parallel.

## Acknowledgments

The authors would like to thank Nicolo Cartiglia (INFN Torino) for providing LGAD data, Paulo Moreira and Szymon Kulis (CERN) for sharing the I/O pads, and Christophe de La Taille and Nathalie Seguin-Moreau (OMEGA Ecole Polytechnique) for insightful discussions.

This work has been authored by Fermi Research Alliance, LLC under Contract No. DE-AC02-07CH11359 with the US Department of Energy, Office of Science, Office of High Energy Physics.